\documentclass{iau}
\usepackage{graphicx}
\usepackage{natbib}
\title[Zooming in on PPD Formation] 
{Zooming in on the Formation of Protoplanetary Disks}

    \author[{\AA}. Nordlund, T. Haugb{\o}lle, M. Kuffmeier, P. Padoan, A. Vasileiades]
    {{\AA}ke Nordlund$^{1,2}$, Troels Haugb{\o}lle$^1$, Michael K{\"u}ffmeier$^{1,2}$, Paolo Padoan$^3$, \and Aris Vasileiades$^{1,2}$}

\affiliation{$^1$Centre for Star and Planet Formation, University of Copenhagen, \\
{\O}ster Voldgade 5, DK-2100 Copenhagen, Denmark \\ email: {\tt aake@nbi.dk} \\[\affilskip]
$^2$Niels Bohr Institute, University of Copenhagen \\
Juliane Maries Vej 30, DK-2100 Copenhagen, Denmark \\[\affilskip]
$^3$ICREA \& ICC, University of Barcelona, Marti i Franquès 1, E-08028 Barcelona, Spain}

\pubyear{2013}
\volume{299}  
\pagerange{xxx-xxx}
 \jname{Exploring the Formation and Evolution of Planetary Systems} \editors{A.C. Editor, B.D. Editor \& C.E. Editor, eds.}
\begin{document}

\def\mspy{\mbox{M}_{\odot}\,\mbox{yr}^{-1}}

\maketitle

\begin{abstract}
We use the adaptive mesh refinement code RAMSES to model the formation of protoplanetary disks in realistic star formation environments. The resolution scales over up to 29 powers of two ($\sim$ 9 orders of magnitude) covering a range from outer scales of 40 pc to inner scales of 0.015 AU. The accretion rate from a 1.5 solar mass envelope peaks near $10^{-4}$ $\mspy$ about 6 kyr after sink particle formation and then decays approximately exponentially, reaching $10^{-6}$ $\mspy$ in 100 kyr. The models suggest universal scalings of physical properties with radius during the main accretion phase, with kinetic and / or magnetic energy in approximate balance with gravitational energy. Efficient accretion is made possible by the braking action of the magnetic field, which nevertheless allows a near-Keplerian disk to grow to a 100 AU size. The magnetic field strength ranges from more than 10 G at 0.1 AU to less than 1 mG at 100 AU, and drives a time dependent bipolar outflow, with a collimated jet and a broader disk wind.

\end{abstract}

\firstsection 
\section{Introduction}
We present first-of-a-kind \textit{ab initio} simulations of the formation of protoplanetary disks, using a locally developed version of the public domain AMR code RAMSES \citep{Ramses1,Ramses2} to zoom in from giant molecular cloud (GMC) scales to the scales of circumstellar disks. The purpose of this work is to characterize the typical properties of accretion onto solar mass protostars with as few free parameters as possible, and (in the near future) to gather a statistical sample of such conditions, to quantify the extent of statistical variation of properties. This is a vast improvement over models where initial and boundary conditions have to be parametrized and a multi-dimensional parameter space has to be sampled. Here the initial and boundary conditions follow instead from the statistical properties of the interstellar medium (ISM), which are reasonably well established, as per for example \cite{Larson2,Solomon+} and \cite{BNrelation}.  We anticipate that since accretion properties are expected to have unavoidable statistical variation from system to system, uncertainties in the properties of the ISM on GMC scales may in fact be of secondary importance. In this contribution we report on the results of a single case study, where a prestellar core in a filament collapses and forms a relatively isolated 1.5 solar mass star. We expect that due to mass loss by bipolar outflows, stars ending up with exactly one solar mass form from systems that accrete more than one solar mass --- indeed, one of the future goals of this line of work is to quantify the amount of mass lost in outflows during the accretion process.

\section{Methods}
The simulations proceed in three steps, with the first step covering 16 levels of (factor two) refinement (smallest cell size $\sim$ 125 AU), following individual star formation in a 40 pc GMC model for $\sim$ 10 Myrs with velocity dispersion and density statistics consistent with Larson's relations \citep{Larson2}, and the kinetic energy maintained by supernova explosions rather than artificial forcing. In the second step, which uses 22 levels of refinement, the neighborhoods of individual stars with a final system mass of 1-2 solar masses are followed during their accretion process, with a smallest mesh size of 2 AU, sufficient to follow the development of the large scale structure of their accretion disks during most of the accretion process. Finally a selection of these disks are studied over time intervals of the order 100 yr at three different epochs, using 29 levels of refinement and cell sizes ranging down to 0.015 AU, sufficient to study and compare the local dynamics of the disks at these different epochs. As is customary for RAMSES simulations, we count levels of refinement compared to the full box size. The root grid is at level 7.

To be able to carry out these simulations we have enhanced the public domain version of the RAMSES code \citep{Ramses1,Ramses2} in several ways. In addition to the MPI parallelization we have added OpenMP parallelization of all routines that contribute significantly to the computing time, thus improving load balancing by allowing a larger number of grids to be handled by each MPI thread. The handling of Hilbert-Peano keys has been modified to allow more levels of refinement to be used. A new sink particle recipe allows the (symplectic) tracking of stellar (sink particle) dynamics, with massive stars ending their lives in supernova explosions after mass dependent life times interpolated from tables. Utilizing a galilean transformed variant of the built-in `geometric refinement', which follows individual sinks, we are able to zoom in to concentric sets of domains around selected sink particles. A combination of adaptive mesh refinement criteria, including Jeans length, and density, pressure and velocity gradients are used, with a roughly constant number of cells per level, resulting in an approximately linear dependence of the computing time on the number of levels.

\section{Results}
Perusal of the accretion histories of different $\sim$ solar mass stars shows that the accretion time (defined for example as the time needed to accrete 90\% of the final mass) can vary from less than 30 kyr to more than 300 kyr, depending on circumstances. Here we report on a case where the 90\% accretion time is about 75 kyr. Figure 1 shows the accretion history and distribution of mass with radius. Within 6 kyr of the sink particle creation the accretion rate reaches a peak at $\approx 6\,10^{-5}$ $\mspy$. The mass of the sink particle at that time is about 10\% of the final mass. Decreasing approximately exponentially with time after that, the accretion rate reaches $10^{-6}$ $\mspy$ after about 100 kyr. As illustrated by the right hand side panel of Fig.\ 1, during the main accretion phase the radial mass distribution follows approximately the same $r^{3/2}$ distribution that one would expect from a free fall situation, with radial velocities scaling as $r^{-1/2}$ and volume densities $\rho\sim r^{-3/2}$.

\begin{figure}[t!]
\begin{center}
\includegraphics[width=6.65cm]{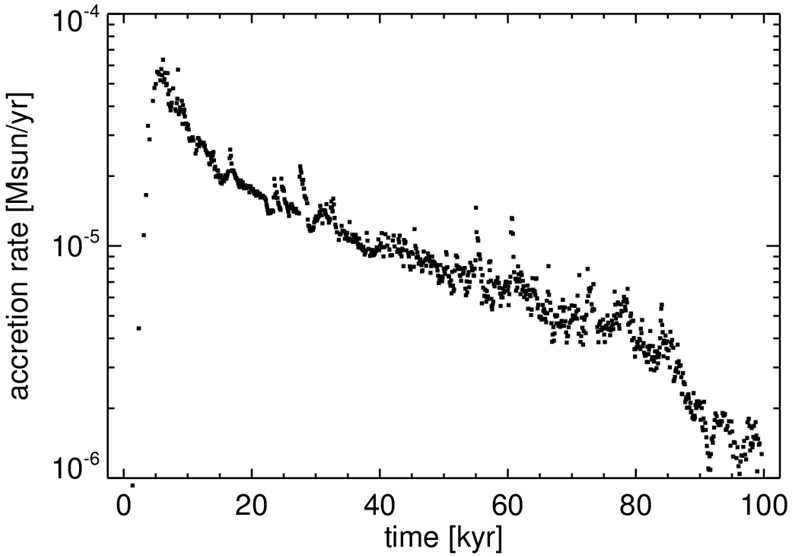}
\includegraphics[width=6.75cm]{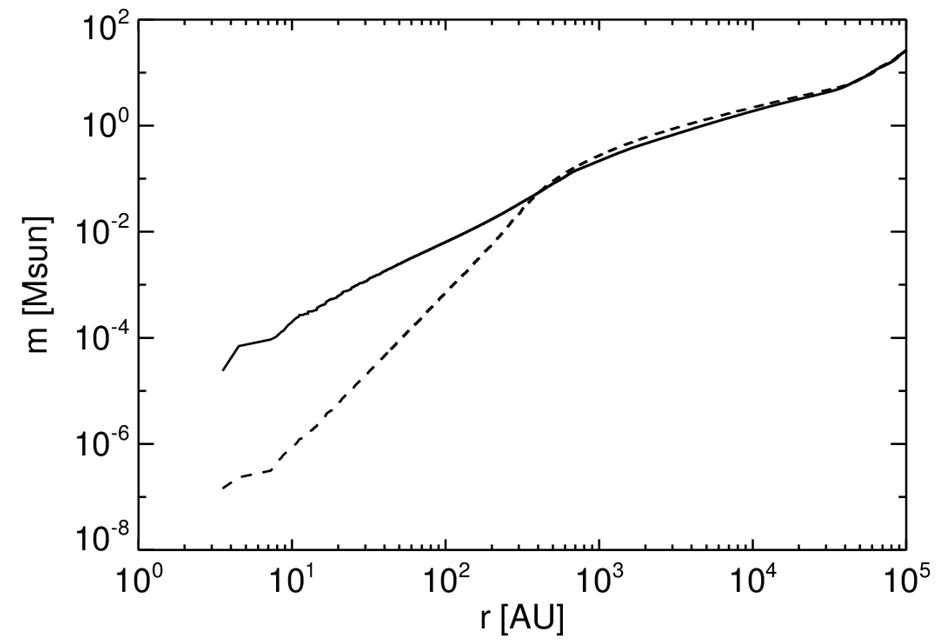}
\end{center}
\caption{Mass accretion rate to the central sink particle as a function of time for the model with 22 levels of refinement (left), and radial profiles of mass, integrated over radii (right), initially (dashed) and during the main accretion phase (full drawn).} \label{fig2}
\end{figure}

\begin{figure}[t!]
\begin{center}
\includegraphics[height=3.03cm]{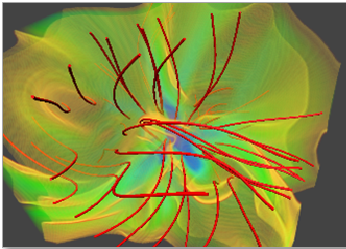}
\includegraphics[height=3.00cm]{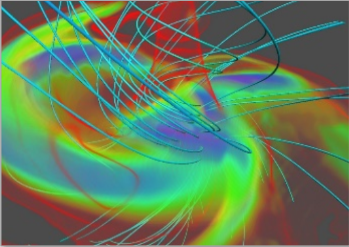}
\includegraphics[height=3.00cm]{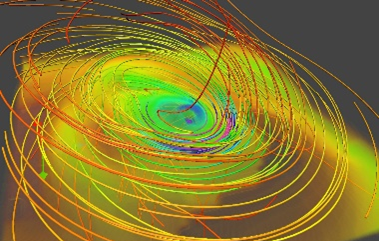}
\end{center}
\caption{Visualizations of the mass density and magnetic field lines, at three epochs after the formation of a sink particle that ended up accreting $\approx1.5$ solar masses: 2 kyr after formation (left), 50 kyr after formation (center), and 100 kyr after formation (right). The point of view is essentially face-on in the left hand side panel.} \label{fig1}
\end{figure}

Three epochs during this evolution were singled out for study at the highest spatial resolution, using 29 levels of refinement; the first one at $t\approx$2 kyr, during the initial phase of rising accretion, the second one at $t\approx$50 kyr, while the accretion rate is still of the order $10^{-5}$ $\mspy$, and the third one at $t\approx$100 kyr, when the accretion rate has dropped to about $10^{-6}$ $\mspy$. Figure 2 shows 3-D VAPOR \citep{Vapor} renderings of the mass density and magnetic field lines at the three epochs, illustrating striking differences. As shown also by Fig.\ 3, the extent of the rotationally supported disk is initially very small--of the order of 1 AU--and the magnetic field lines are hour-glass shaped. At the second epoch the extent of the rotationally supported disk is $\approx 10$ AU, the magnetic field lines are wound into spiral shapes, supporting a central, well-collimated (but essentially one-sided) jet and a broader disk wind outflow. Finally, during the third epoch, the rotationally supported disk extends out to $\approx 100$ AU, the magnetic field lines are wound much tighter than in the 2nd case, and there is only a weak disk wind outflow.

Figure 3 shows the (mass weighted) average speed of rotation in shells of constant radius, as well as the Alfv{\'e}n speed (based on the rms magnetic field and the average mass density), and the velocity dispersion relative to the average speed of rotation. The line with constant slope shows the final Keplerian speeds of rotation, illustrating the near-equipartion that apparently exist between gravitational and kinetic energy at small radii and between gravitational and magnetic energy at large radii.

\section{Discussions and Conclusions}
One of the most important results of this work is that it illustrates the self-consistent development of a situation that satisfies two of the critical requirements necessary to form realistic protoplanetary disks: 1) The initial very efficient shedding of angular momentum, allowing half of the stellar mass to be accreted in less than 25 kyr, while 2) nevertheless allowing an extended rotationally supported protoplanetary disk to develop and grow in size towards the end of the accretion phase. The presence of a strong azimuthal magnetic field in the disk could suppress shear driven turbulence and allow the \citet{GoldreichWard} mechanism to operate.

At larger radii than those where a rotationally supported disk exists there is instead approximate balance between gravitational and magnetic energy density, while the kinetic energy density is much smaller.  This may be seen as a balance between gravity trying to pull in and compress the magnetic field and magnetic forces resisting the pull. This is supported by visualizations (available as part of the on-line copy of the symposium presentation) illustrating the chaotic nature of the mass density and magnetic field, extending over scales with widely varying time scales (essentially following a free-fall time scaling $t_{\mbox{ff}}\sim \rho^{-1/2}$). The widely different time scales at different radii presumably make an ordered collapse impossible, resulting instead in a generic, chaotic dynamical state, which allows accretion to proceed at some significant fraction of the free fall speed.

\begin{figure}[t!]
\begin{center}
\includegraphics[width=7.5  cm]{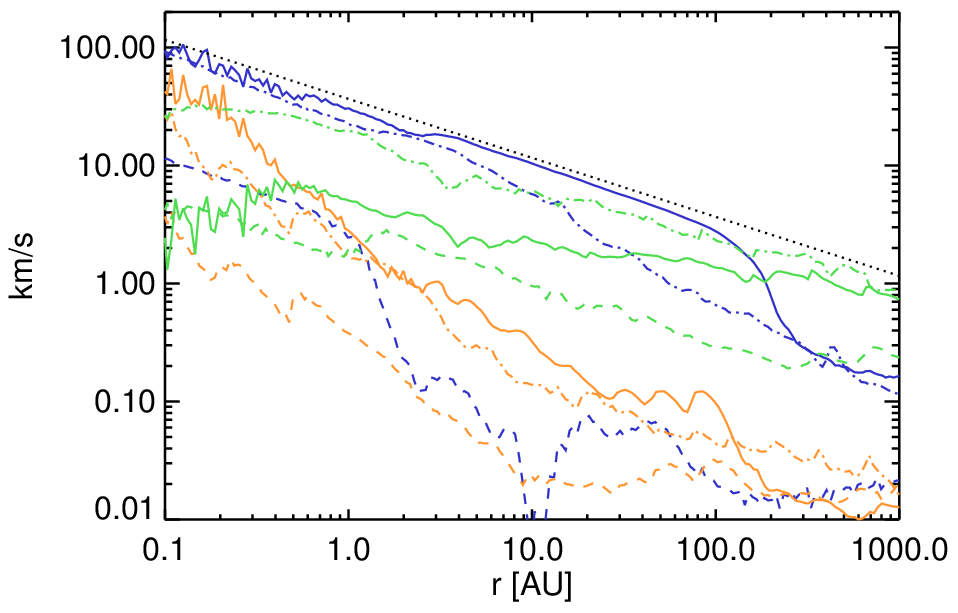}
\end{center}
\caption{Radial profiles of characteristic velocities at three epochs; 2 kyr (dashed), 50 kyr (dot-dashed), and 100 kyr (full drawn). From top to bottom the three curves (blue, green and orange in the on-line version) for each epoch are: the mass weighted rotational velocity, the average Alfv{\'e}n speed (see the text), and the root mean square velocity dispersion. The dotted straight line (slope -1/2) shows the Keplerian rotation speed for the fully accreted star.} \label{fig3}
\end{figure}

As a byproduct of this type of modeling, which starts out from a supernova driven interstellar medium, we can follow the transport of short-lived radionuclides (SLRs), from the time of ejection from supernovae and until they become part of the proto-planetary disks.  As shown by \cite{Aris13} the transport time is on average short enough to be consistent with initial abundance of 26Al in the Solar System derived from cosmochemistry. Of particular interest in the near future is to characterize the amount of variation with time of the SLR abundance during the life time of PP-disks surrounding solar mass stars.

\vspace{6pt}\noindent \textbf{Acknowledgements:} Research at Centre for Star and Planet Formation is funded by the Danish National Research Foundation. PP is supported by the FP7-PEOPLE-2010-RG grant PIRG07-GA-2010- 261359. Supercomputing resources at  Leibniz Rechenzentrum (PRACE project pr86li) and at DeIC/KU in Copenhagen are gratefully acknowledged.




\end{document}